\documentclass[reqno,11pt]{article}
\usepackage{amssymb,amsthm,amsmath,amsfonts}
\usepackage{epsfig}

\theoremstyle{plain}
\begingroup
 
\newtheorem{thm}{Theorem}[section] 

\newtheorem{lemma}[thm]{Lemma}

\endgroup
\theoremstyle{definition}

\theoremstyle{remark}

\numberwithin{equation}{section}
\numberwithin{figure}{section}

\begin{document}
\title{Large-time rescaling behaviors of Stokes and Hele-Shaw  flows driven by injection}
\author{Yu-Lin Lin\textsuperscript{1}}

\date{\today}

\maketitle

\begin{abstract}
In this paper, we give a precise description of  the rescaling behaviors of  global strong polynomial solutions to the reformulation of zero surface tension Hele-Shaw problem driven by  injection, the Polubarinova-Galin equation, in terms of Richardson complex moments. From past results,  we know that this set of solutions is large. This method can also be applied to zero surface tension Stokes flow driven by injection and a rescaling behavior is given in terms of many conserved quantities as well. \end{abstract} 


\footnotetext[1]
{Institute of Mathematics, Academia Sinica, Nankang, Taipei, Taiwan 11529 R.O.C. .\\
Email: \tt{yulin@math.brown.edu}}

 \section{Introduction}
\label{sec1}
This present paper is mainly devoted to two problems: two-dimensional zero surface tension Hele-Shaw flows driven by injection at the origin and two-dimensional zero surface tension Stokes flow driven by injection at the origin. In both cases, ingenious complex variable methods have been developed which use conformal mapping. Let $\Omega(t)$ be the moving domain and  $f(\xi,t):D\rightarrow \Omega(t)$ is univalent and analytic in $\overline{D}$, $f(0,t)=0$, $f^{'}(0,t)>0$ . We set 
\[\frac{\partial}{\partial t}f(\xi,t)=f_{t}(\xi,t),\quad\frac{\partial}{\partial\xi}f(\xi,t)=f^{'}(\xi,t),\quad D=\{x||x|<1\}.\]
For the former problem, the 
differential equation with injection strength $2\pi$ at the origin is, 
\begin{equation}
\label{PG1}
Re\left[f_{t}(\xi,t)\overline{f^{'}(\xi,t)\xi}\right]=1, \quad\xi\in \partial D
\end{equation}
Equation $(\ref{PG1})$ is called the Polubarinova-Galin equation since Galin and Polubarinova-Kochina first derived it and investigated the Riemann mapping method along these lines. A solution to equation $(\ref{PG1})$ is said to be a strong solution for $t\in [0,b)$ if $f(\xi,t)$ is univalent and analytic in
$\overline{D}$, $f(0,t)=0$, $f^{'}(\xi,0)>0$ and continously
differentiable in $t$, $t\in [0,b)$.  For the later problem, the differential equation with injection strength $2\pi$ at the origin is,
\begin{equation}
\label{stokes1}
\frac{\partial}{\partial t}\left[f'(\xi,t)\overline{f}(\frac{1}{\xi},t)\right]+2\Lambda'(\xi,t)=0, \quad\xi\in\partial D
\end{equation}
where $\Lambda(\xi,t)=\chi(f(\xi,t))$ and $\chi(z)\sim -\log z$ as $z\rightarrow 0$.\par
Define
\[O(E)=\left\{f\mid \mbox{$f$ is univalent and analytic in $E$}, f(0)=0
,f^{'}(0)>0 \right\}.\] 
In the case of ZST Hele-Shaw flows driven by injection, it is shown in Gustafsson~\cite{gustaf1} that starting with a degree $n$ polynomial mapping $f_{n}(\xi,0)\in O(\overline{D})$, the strong solution to (\ref{PG1}) $f_{n}(\xi,t)$ is also a polynomial of the same degree. In Huntingford~\cite{hford}, it is shown that not any given polynomial function $f(\xi,0)\in O(\overline{D})$ can produce a global strong polynomial solution.  However, it is proven in Gustafsson, Prokhorov and Vasil'ev~\cite{gustaf2} that starting with a starlike mapping $f(\xi,0)\in O(\overline{D})$, the strong solution to (\ref{PG1}) $f(\xi,t)$ is global. In the case of ZST Stokes flow driven by injection, in Cummings, Howison and King~\cite{cummings}, the authors obtain many polynomial solutions to (\ref{stokes1}). The global dynamics is not as clear as that of Hele-Shaw flows but some of polynomial solutions to (\ref{stokes1}) are global.\par
In this paper, we focus on large-time rescaling behaviors of  global polynomial solutions to ZST Hele-Shw flows driven by injection and ZST Stokes flow driven by injection in terms of  many invariant quantities. These invariant quantities are defined in each problem respectively as follows. Let a family of domains $\{\Omega(t)\}_{t\geq 0}$ be a solution to ZST Hele-Shaw flows driven by injection with strength $2\pi$ or ZST Stokes flow driven by injection with strength $2\pi$ and $f(\xi,t)\in O(\overline{D})$ be the corresponding Riemann mapping. In the former problem, Richardson complex moments are defined as
\[M_{k}(t)=\frac{1}{\pi}\int_{\Omega(t)}z^{k}dxdy=\frac{1}{2\pi i}\int_{|\xi|=1}f^{k}(\xi,t)f'(\xi,t)\overline{f}(\frac{1}{\xi},t)d\xi,\quad z=x+iy.\]
The rate of change of moments is given by
\begin{equation}
\label{11}
\frac{d}{dt}M_{k}(t)=2\delta_{k0},\notag\\
\end{equation}
where $\delta_{k0}=1$ when $k=0$ and $\delta_{k0} =0$ if $k \neq 0$.  In particular, the moments $M_{k},$ $k\geq 1$ are conserved. The zero moment $M_{0}(t)$ equals $\frac{1}{\pi}\mid\Omega(t)\mid$ where $\mid\Omega(t)\mid$ is the area of the domain $\Omega(t)$.  In the later problem, it is shown in Cummings, Howison and King~\cite{cummings} that the quantities $C_{k}(t), k\geq 0$ where 
\[C_{k}(t)=\frac{1}{2\pi i}\int_{|\xi|=1}\xi^{k}f'(\xi,t)\overline{f}(\frac{1}{\xi},t)d\xi\]
satisfy
\[\frac{d}{dt}C_{k}(t)=2\delta_{k0}.\]
In particular, the quantities, $C_{k}, k\geq 1$ are conserved. The quantity $C_{0}(t)=C_{0}(0)+2t$ equals $\frac{1}{\pi}\mid\Omega(t)\mid$.
\par
This current paper is organized as follows.  In section~\ref{sec4}, we assume $f_{n}(\xi,t)=\sum_{i=1}^{n}a_{i}(t)\xi^{i}$ a global strong degree $n\geq 2$ polynomial solution to (\ref{PG1}) and obtain how each coefficient decays or grows in terms of Richardson complex moments. By this result, we obtain many rescaling behaviors for $f_{n}(\xi,t)$. Let $n_{0}=\min\{k\geq 1|M_{k}\neq 0\}$, then it is  shown that 
\[\lim_{t\rightarrow\infty}\left[f_{n}(\xi,t)-\sqrt{2t+M_{0}(0)}\xi\right](2t)^{\frac{n_{0}+1}{2}}=\overline{M_{n_{0}}}\xi^{n_{0}+1}\neq 0\]  
and 
\[\lim_{t\rightarrow\infty}\left[f_{n}(\sqrt{2t}\xi,t)-\left(2t+\frac{M_{0}(0)}{2}\right)\xi-\sum_{k=2}^{n}\overline{M_{k-1}}\xi^k\right](\sqrt{2t})^2\]\[=\frac{-M_{0}^2(0)}{8}\xi-\frac{M_{0}(0)}{2}\sum_{k=2}^{n}k\overline{M_{k-1}}\xi^{k}\neq 0.\]
Thus, the rescaled limit depends on the initial data only through the conserved moments. A geometric characterization of a rescaling behavior for the corresponding moving domain $\Omega(t)=f_{n}(D,t)$ is given as well. By rescaling $\Omega(t)$ to be $\Omega'(t)$ which has area  $\pi$ only, we obtain that the radius and curvature of $\Omega'(t)$ decay to $1$ algebraically as $t^{-(n_{0}/2+1)}$.  In section~\ref{sec8}, it is shown that  for a global degree $n\geq 2$ polynomial solution $f_{n}(\xi,t)=\sum_{i=1}^{n}a_{i}(t)\xi^{i}$ to (\ref{stokes1}), 
\[\lim_{t\rightarrow\infty}\left[f_{n}(\xi,t)-\sqrt{2t+C_{0}(0)}\xi\right]\sqrt{2t}=\sum_{k=2}^{n}\overline{C_{k-1}}\xi^{k}\neq 0.\]
A geometric characterization of a rescaling behavior for the corresponding moving domain $\Omega(t)=f_{n}(D,t)$ is given as well. By rescaling $\Omega(t)$ to be $\Omega'(t)$ which has area  $\pi$ only, we obtain that the radius and curvature of $\Omega'(t)$ decay to $1$ algebraically as $t^{-1}$.


\section{Rescaling behavior in Hele-Shaw flows}
\label{sec4}
If we assume that $f_{n}(\xi,t)=\sum_{i=1}^{n}a_{i}\xi^{i}$ is a  strong degree $n$ polynomial solution to the P-G equation $(\ref{PG1})$, and that $\Omega(t)=f_{n}(D,t)$, then by the result in Richardson~\cite{richardson}, the moments $\{M_{k}(t)\}_{0\leq k\leq n-1}$ can be represented by these coefficients $\{a_{k}\}_{k\geq 1}$ as
\begin{equation}
\label{representation}
M_{k}(t)= \sum_{i_{1},\cdots,i_{k+1}}i_{1}a_{i_{1}}a_{i_{2}}\cdots a_{i_{k+1}}\overline{a_{i_{1}+\cdots+i_{k+1}}}.
\end{equation}
The zero moment $M_{0}(t)=\sum_{i=1}^{n}i\mid a_{i}(t)\mid^2=2t+M_{0}(0)=a_{1}^{2}(t)+g(t)$ where $g(t)=\sum_{i=2}^{n}i|a_{i}(t)|^2$. In the case that $f_{n}(\xi,t)$ is global, by Kuznetsova~\cite{kuz}, $a_{1}^2(t)-2t$ is nondecreasing and $g(t)$ is not increasing. Therefore, the coefficient $a_{1}(t)$ satisfies 
\[\lim_{t\rightarrow\infty}\frac{a_{1}(t)}{\sqrt{2t}}=1\]
and $a_{k}(t), k\geq 2$ is bounded by $\sqrt{g(0)}$ in this case.\par
Now given a real-coefficient  global strong solution $a_{1}(t)\xi+a_{2}(t)\xi^2+a_{3}(t)\xi^3$ to (\ref{PG1}), then by (\ref{representation}), $a_{2}(t)$ and $a_{3}(t)$ are expressed in terms of  $a_{1}(t)$ and the conserved moments $\{M_{k}\}_{1\leq k\leq 2}$ as follows:
\[a_{2}(t)=M_{1}\frac{a_{1}^2(t)}{3M_{2}+a_{1}^4(t)}\]
and
\[a_{3}(t)=\frac{M_{2}}{a_{1}^3(t)}.\]
The moments in this case are real-valued as well since $\{a_{k}(t)\}_{1\leq k\leq 3}$ are real-valued. We can see that the coefficients $\{a_{k}(t)\}_{2\leq k\leq 3}$ in this example satisfy
\[\lim_{t\rightarrow\infty}a_{2}(t)(a_{1}(t))^2=M_{1}=\overline{M_{1}}\]
and
\[\lim_{t\rightarrow\infty}a_{3}(t)(a_{1}(t))^3=M_{2}=\overline{M_{2}}.\]
Moreover, 
\[a_{2}(t)a_{1}^2(t)-\overline{M_{1}}=a_{2}(t)a_{1}^2(t)-M_{1}=-\frac{3M_{1}M_{2}}{3M_{2}+a_{1}^4(t)}=O\left(\frac{1}{a_{1}^4}\right),\]
\[a_{3}(t)a_{1}^3(t)-\overline{M_{2}}=a_{3}(t)a_{1}^3(t)-M_{2}=0.\]
\par
We give a general statement of this kind of decay for coefficients of global strong polynomial solutions to (\ref{PG1}) in subsection~\ref{sub4.1}, and use the results to obtain rescaling behaviors of  global strong polynomial solutions in subsection~\ref{sub4.2}. If the lower Richardson moments vanish, the decay rate of coefficients is much faster and the global strong polynomial solutions to (\ref{PG1}) have better rescaling behaviors. In subsection~\ref{sub4.3}, a geometric characterization of a rescaling behavior for the moving domain is given. 

\subsection{Decay of coefficients in terms of moments}
\label{sub4.1}
\begin{lemma}
\label{Lemma4.1}
If $f_{n}(\xi,t)=\sum_{i=1}^{n} a_{i}(t)\xi^{i}$ is a global strong polynomial solution  of degree $n\geq 2$ to the P-G equation $(\ref{PG1})$, then for $k\geq 2$
\begin{equation}
\label{decayrate}
\lim_{t\rightarrow\infty}a_{1}^ka_{k}=\overline{M_{k-1}}.
\end{equation}
Furthermore, for $k\geq 2$,
\begin{equation}
\label{199}
a_{1}^{k}a_{k}-\overline{M_{k-1}}=O\left(\frac{1}{a_{1}^4(t)}\right).
\end{equation}
\end{lemma}
\begin{proof}We want to prove (\ref{decayrate}) by induction.\\
Step1: Since $a_{n}=\overline{M_{n-1}}a_{1}^{-n}$, it is clear that $(\ref{decayrate})$ holds for $k=n$.\\
Step2: Assume that the (\ref{decayrate}) holds for $n-s_{0}\leq  k\leq n$ where $s_{0}+1\leq n-2$. \\
Claim: The result $(\ref{decayrate})$ holds for $k=n-(s_{0}+1)$.\begin{proof}{(of claim)}
For $k=n-(s_{0}+1)$,
\begin{align}
         M_{k-1}(t)=&\sum_{i_{1},\cdots,i_{k}}i_{1}a_{i_{1}}a_{i_{2}}\cdots a_{i_{k}}\overline{a_{i_{1}+\cdots+i_{k}}}\notag\\
=&a_{1}^{k}\overline{a_{k}}+\sum_{i_{1},\cdots,i_{k};\prod_{j=1}^{k}i_{j}\not=1}i_{1}a_{i_{1}}a_{i_{2}}\cdots a_{i_{k}}\overline{a_{i_{1}+\cdots+i_{k}}}\notag
 \end{align}
If $\prod_{j=1}^{k}i_{j}\neq 1$, then $a_{i_{1}+\cdots+i_{k}}=O(\frac{1}{a_{1}^{i_{1}+\cdots+i_{k}}})$ since $i_{1}+\cdots+i_{k}\geq n-s_{0}$ and we assume that (\ref{decayrate}) holds for $n-s_{0}\leq k\leq n$. By the above and that  $|a_{i}|\leq \sqrt{g(0)}, i\geq 2$,  then $|a_{i_{1}}a_{i_{2}}\cdots a_{i_{k}}\overline{a_{i_{1}+\cdots+i_{k}}}|=O(\frac{1}{a_{1}^{2}})$ and therefore
 \[M_{k-1}(t)=a_{1}^{k}\overline{a_{k}}+O\left(\frac{1}{a_{1}^2}\right).\]
Therefore,
\[\lim_{t\rightarrow\infty}a_{1}^k\overline{a_{k}}=M_{k-1}.\]
By taking complex conjugate of the above, we obtain (\ref{decayrate}).
\end{proof}
Furthermore, since (\ref{decayrate}) and that  $|a_{i}|\leq \sqrt{g(0)}, i\geq 2$, then $|a_{i_{1}}a_{i_{2}}\cdots a_{i_{k}}\overline{a_{i_{1}+\cdots+i_{k}}}|=O(\frac{1}{a_{1}^4})$ and therefore
\[a_{1}^{k}\overline{a_{k}}-M_{k-1}=-\sum_{i_{1},\cdots,i_{k};\prod_{j=1}^{k}i_{j}\not=1 } i_{1}a_{i_{1}}a_{i_{2}}\cdots a_{i_{k}}\overline{a_{i_{1}+\cdots+i_{k}}}=O\left(\frac{1}{a_{1}^4}\right).\]
By taking complex conjugate of the above, we obtain (\ref{199}).

\end{proof}
In the case that the lower Richardson moments vanish, better qualitative properties for the coefficients $a_{1}(t)$ and $\{a_{k}(t)\}_{k\geq 2}$ are given as follows.
\begin{lemma}
\label{lemma4.2}
Let $f_{n}(\xi,t)=\sum_{i=1}^{n}a_{i}(t)\xi^{i}$ be a global strong degree $n\geq 2$ polynomial solution to $(\ref{PG1})$. Denote $n_{0}=\min\{k\geq 1\mid M_{k}\not=0\}$. Then\\
(a)
\begin{equation}
\label{moment5}
\lim_{t\rightarrow\infty}a_{1}^{n_{0}+1}a_{n_{0}+1}=\overline{M_{n_{0}}},
\end{equation}
\begin{equation}
\label{moment6}
\lim_{t\rightarrow\infty}a_{1}^{k}a_{k}=\overline{M_{k-1}},\quad k>n_{0}+1,
\end{equation}
and
\begin{equation}
\label{n_{0}}
\lim_{t\rightarrow\infty}a_{1}^{n_{0}+1}a_{k}=0,\quad 2\leq k< n_{0}+1.
\end{equation}
(b)\begin{equation}
\label{first}
\lim_{t\rightarrow\infty}\left[a_{1}(t)-\sqrt{2t+M_{0}(0)}\right](2t)^{n_{0}+3/2}=\frac{-(n_{0}+1)}{2}\left|M_{n_{0}}\right|^2.
\end{equation}
\end{lemma}
\begin{proof}(a)
If $n_{0}=1$, part (a) in Lemma~\ref{lemma4.2} follows from Lemma~\ref{Lemma4.1}. Now assume $n_{0}\geq 2$. We split the proof for $n_{0}\geq 2$ into two parts.\par
{\bf claim1:}
\[
\lim_{t\rightarrow\infty}a_{1}^{n_{0}+1}a_{n_{0}+1}=\overline{M_{n_{0}}},
\]
and 
\[
\lim_{t\rightarrow\infty}a_{1}^{k}a_{k}=\overline{M_{k-1}},\quad k>n_{0}+1.
\]
\begin{proof}{(of claim1)} As shown in $(\ref{decayrate})$.\\
\end{proof}

{\bf claim2:}
\[
\lim_{t\rightarrow\infty}a_{1}^{n_{0}+1}a_{k}=0,\quad 2\leq k\leq n_{0}.
\]
\begin{proof}{(of claim2)}
In $(\ref{representation})$, the Richardson moments have the following representation:
\begin{align}
M_{k}(t)=&\sum_{i_{1},\cdots,i_{k+1}}i_{1}a_{i_{1}}a_{i_{2}}\cdots a_{i_{k+1}}\overline{a_{i_{1}+\cdots+i_{k+1}}}\notag\\
=&a_{1}^{k+1}\overline{a_{k+1}}+\sum_{i_{1},\cdots,i_{k+1};\prod_{j=1}^{k+1}i_{j}\not=1}i_{1}a_{i_{1}}a_{i_{2}}\cdots a_{i_{k+1}}\overline{a_{i_{1}+\cdots+i_{k+1}}}\notag.
\end{align}
This means
\begin{equation}
\label{moment0}
\overline{a_{k+1}}=\frac{1}{a_{1}^{k+1}}\left[M_{k}-\sum_{i_{1},\cdots,i_{k+1};\prod_{j=1}^{k+1}i_{j}\not=1}i_{1}a_{i_{1}}a_{i_{2}}\cdots a_{i_{k+1}}\overline{a_{i_{1}+\cdots+i_{k+1}}}\right].
\end{equation}
By taking complex conjugate, the above becomes
\begin{equation}
\label{moment3}
a_{k+1}=\frac{1}{a_{1}^{k+1}}\left[\overline{M_{k}}-\sum_{i_{1},\cdots,i_{k+1};\prod_{j=1}^{k+1}i_{j}\not=1}i_{1}\overline{a_{i_{1}}a_{i_{2}}\cdots a_{i_{k+1}}}a_{i_{1}+\cdots+i_{k+1}}\right].
\end{equation}
Hence
\begin{equation}
\label{moment4}
a_{k+1}=\frac{1}{a_{1}^{k+1}}\left[\overline{M_{k}}-O\left(\frac{1}{a_{1}^4}\right)\right].
\end{equation}
By substituting  $n_{0}-1$ for $k$ in $(\ref{moment4})$ and applying the fact that $M_{n_{0}-1}=0$, we can first show that 
\[\lim_{t\rightarrow\infty}a_{1}^{n_{0}+1}a_{n_{0}}=0.\]
If $n_{0}=2$, we are done. Assume $n_{0}\geq 3$ now and continue the proof by induction. \par
Assume for some $s_{0}$ where $0\leq s_{0}\leq n_{0}-3$, 
\begin{equation}
\label{induction}
\lim_{t\rightarrow\infty}a_{1}^{n_{0}+1}a_{n_{0}-s}=0,\quad 0\leq s\leq s_{0}.
\end{equation}
We need to show that (\ref{induction}) holds for $s=s_{0}+1$. Denote $l=n_{0}-(s_{0}+1)$. In $(\ref{moment3})$, by substituting $l$ for $k+1$, we have that 
\begin{equation}
\label{p1}
a_{l}=\frac{-1}{a_{1}^{l}}\left[\sum_{i_{1},\cdots,i_{l};\prod_{j=1}^{l}i_{j}\not=1}i_{1}\overline{a_{i_{1}}a_{i_{2}}\cdots a_{i_{l}}}a_{i_{1}+\cdots+i_{l}}\right].
\end{equation}
However, if $\prod_{j=1}^{l}i_{j}\not=1,$ then $i_{1}+\cdots+i_{l}\geq l+1$. Therefore, $a_{i_{1}+\cdots+i_{l}}=O(\frac{1}{a_{1}^{n_{0}+1}})$ due to  the assumption in $(\ref{induction})$, results $(\ref{moment5})$ and  $(\ref{moment6})$. Hence 
\begin{equation}
\label{p2}
i_{1}\overline{a_{i_{1}}a_{i_{2}}\cdots a_{i_{l}}}a_{i_{1}+\cdots+i_{l}}=a_{1}^{l-1}O\left(\frac{1}{a_{1}^{n_{0}+1}}\right).
\end{equation}
Finally, by  the representation (\ref{p1}) for $a_{l}$ and (\ref{p2}), we obtain the estimate
\[ a_{l}=\frac{-1}{a_{1}^{l}}\left[a_{1}^{l-1}O\left(\frac{1}{a_{1}^{n_{0}+1}}\right)\right]=O\left(\frac{1}{a_{1}^{n_{0}+2}}\right)\]
and this implies
\[\lim_{t\rightarrow\infty}a_{l}a_{1}^{n_{0}+1}=0.\]
This says (\ref{induction}) also holds for $s=s_{0}+1$.\par
Hence, claim2 is proven by induction.
\end{proof}
By claim1 and claim2, the proof for part (a) in Lemma~\ref{lemma4.2} is done.\\
(b)\begin{align}
&a_{1}(t)-\sqrt{M_{0}(0)+2t}\notag\\
 =&\frac{a_{1}^2(t)-M_{0}(0)-2t}{a_{1}(t)+\sqrt{M_{0}(0)+2t}}\notag\\
 =&\frac{-g(t)}{a_{1}(t)+\sqrt{M_{0}(0)+2t}}.\notag
\end{align}
By (\ref{moment5}), (\ref{moment6}) and (\ref{n_{0}}), we can get that
\[\lim_{t\rightarrow}g(t)(2t)^{n_{0}+1}=(n_{0}+1)\left|M_{n_{0}}\right|^2.\]
Therefore, 
\[\lim_{t\rightarrow}\left[a_{1}(t)-\sqrt{M_{0}(0)+2t}\right](2t)^{n_{0}+3/2}=\frac{-(n_{0}+1)}{2}\left|M_{n_{0}}\right|^2.\] 
\end{proof}
\subsection{Rescaling behaviors of global strong polynomial solutions}
\label{sub4.2}
\begin{thm}
\label{thm4.3}
Let $f_{n}(\xi,t)$ be a global strong degree $n\geq 2$ polynomial solution to $(\ref{PG1})$. Denote $n_{0}=\min\{k\geq 1\mid M_{k}\not=0\}$. We can see the following rescaling behaviors:\\
(i)
\begin{equation}
\label{rescaling100}
\lim_{t\rightarrow\infty}\left[ f_{n}(\xi,t)-\sqrt{M_{0}(0)+2t}\xi\right](2t)^{\frac{n_{0}+1}{2}}=\overline{M_{n_{0}}}\xi^{n_{0}+1}.
\end{equation}
(ii)
\begin{align}
\lim_{t\rightarrow\infty}&\left[f_{n}(\sqrt{2t}\xi,t)-\left(2t+\frac{M_{0}(0)}{2}\right)\xi-\sum_{k=2}^{n}\overline{M_{k-1}}\xi^k\right](\sqrt{2t})^2\notag\\
=&\frac{-M_{0}^2(0)}{8}\xi-\frac{M_{0}(0)}{2}\sum_{k=2}^{n}k\overline{M_{k-1}}\xi^{k}.\notag\\
\end{align}
\end{thm}
\begin{proof} 
Denote $f_{n}(\xi,t)=\sum_{i=1}^{n}a_{i}(t)\xi^{i}$.\\
(i)
\begin{align}
&f_{n}(\xi,t)-\sqrt{M_{0}(0)+2t}\xi\notag\\
=&\left(a_{1}(t)-\sqrt{M_{0}(0)+2t}\right)\xi+\sum_{i=2}^{n}a_{i}\xi^{i}\notag\\
=&I+II\notag
\end{align}
where $I=(a_{1}(t)-\sqrt{M_{0}(0)+2t})\xi$ and $II=\sum_{i=2}^{n}a_{i}\xi^{i}$.\par
By  $(\ref{moment5})$, $(\ref{moment6})$ and $(\ref{n_{0}})$ in Lemma~\ref{lemma4.2}, 
\begin{equation}
\label{moment1}
\lim_{t\rightarrow\infty}(2t)^{\frac{1+n_{0}}{2}}(II)=\overline{M_{n_{0}}}\xi^{n_{0}+1}.
\end{equation}
By  (\ref{first}), we can obtain
\begin{equation}
\label{moment7}
\lim_{t\rightarrow\infty}(2t)^{n_{0}+1}(I)=0.
\end{equation}
By (\ref{moment1}) and (\ref{moment7}), we prove (i).\\
(ii)
\[f_{n}(\sqrt{2t}\xi,t)=\sum_{i=1}^{n}a_{i}(t)\left(\sqrt{2t}\right)^{i}\xi^{i}.\]
\[\left[f_{n}(\sqrt{2t}\xi,t)-\left(2t+\frac{M_{0}(0)}{2}\right)\xi-\sum_{k=2}^{n}\overline{M_{k-1}}\xi^k\right]\]
\[=\left[a_{1}(t)\sqrt{2t}-\left(\sqrt{2t}\right)^2-\frac{M_{0}(0)}{2}\right]\xi+\sum_{k=2}^{n}\left[a_{k}(t)\left(\sqrt{2t}\right)^{k}-\overline{M_{k-1}}\right]\xi^{k}.\]
In order to prove (ii), it is enough to show the following claim.\\
{\bf Claim}: For $k\geq 2$
\begin{equation}
\label{moment8}
\lim_{t\rightarrow\infty}\left[a_{k}\left(\sqrt{2t}\right)^{k}-\overline{M_{k-1}}\right]\left(\sqrt{2t}\right)^2=-\frac{M_{0}(0)}{2}(k)\overline{M_{k-1}}
\end{equation}
and
\begin{equation}
\label{moment9}
\lim_{t\rightarrow\infty}\left[a_{1}(t)\sqrt{2t}-\left(\sqrt{2t}\right)^2-\frac{M_{0}(0)}{2}\right](2t)=\frac{-M_{0}^2(0)}{8}.
\end{equation}
\begin{proof}{(of claim)} The proof for $(\ref{moment9})$ is easy. We will now focus on the proof of $(\ref{moment8})$. Equation $(\ref{moment0})$ states
\[
\overline{a_{k}}=\frac{1}{a_{1}^{k}}\left[M_{k-1}-\sum_{i_{1},\cdots,i_{k};\prod_{j=1}^{k}i_{j}\not=1}i_{1}a_{i_{1}}a_{i_{2}}\cdots a_{i_{k}}\overline{a_{i_{1}+\cdots+i_{k}}}\right].
\]
We multiply the above identity  by $(\sqrt{2t})^{k}$ and obtain
\[\label{moment2}\overline{a_{k}}\left(\sqrt{2t}\right)^{k}=\left(\frac{\sqrt{2t}}{a_{1}}\right)^{k}\left[M_{k-1}-\sum_{i_{1},\cdots,i_{k};\prod_{j=1}^{k}i_{j}\not=1}i_{1}a_{i_{1}}a_{i_{2}}\cdots a_{i_{k}}\overline{a_{i_{1}+\cdots+i_{k}}}\right].\]
By subtracting $M_{k-1}$ from the above identity, we have
\begin{align}
\label{timerescaling}
              &\left[\overline{a_{k}}\big(\sqrt{2t}\big)^{k}-M_{k-1}\right]\notag\\
=&M_{k-1}\left[\left(\frac{\sqrt{2t}}{a_{1}}\right)^{k}-1\right]-\left[\sum_{i_{1},\cdots,i_{k};\prod_{j=1}^{k}i_{j}\not=1}i_{1}a_{i_{1}}a_{i_{2}}\cdots a_{i_{k}}\overline{a_{i_{1}+\cdots+i_{k}}}\right]
\notag\\
&-\left[\left(\frac{\sqrt{2t}}{a_{1}}\right)^{k}-1\right]\left[\sum_{i_{1},\cdots,i_{k};\prod_{j=1}^{k}i_{j}\not=1}i_{1}a_{i_{1}}a_{i_{2}}\cdots a_{i_{k}}\overline{a_{i_{1}+\cdots+i_{k}}}\right].
\end{align}
In order to estimate the right-hand side of $(\ref{timerescaling})$, we estimate
\[\left(\frac{\sqrt{2t}}{a_{1}}\right)^{k}-1\mbox{\quad and \quad}
\sum_{i_{1},\cdots,i_{k};\prod_{j=1}^{k}i_{j}\not=1}i_{1}a_{i_{1}}a_{i_{2}}\cdots a_{i_{k}}\overline{a_{i_{1}+\cdots+i_{k}}}\]
in (a) and (b) as follows:\\
(a)\\
By the mean value theorem, there exists $\theta_{k}(t)\in [0,1]$ such that
\[\left(\frac{\sqrt{2t}}{a_{1}}\right)^{k}-1=k\left[1+\theta_{k}(t)\left(\frac{\sqrt{2t}}{a_{1}}-1\right)\right]^{k-1}\left(\frac{\sqrt{2t}}{a_{1}}-1\right).\]
For this term $\frac{\sqrt{2t}}{a_{1}}-1$, we have
\[\frac{\sqrt{2t}}{a_{1}}-1=\frac{\sqrt{2t}-a_{1}}{a_{1}}=\frac{1}{a_{1}}\frac{2t-a_{1}^2}{\sqrt{2t}+a_{1}}=\frac{1}{a_{1}}\frac{g(t)-M_{0}(0)}{\sqrt{2t}+a_{1}}.\]
Therefore,
\[\left(\frac{\sqrt{2t}}{a_{1}}\right)^{k}-1=k\left[1+\theta_{k}(t)\left(\frac{\sqrt{2t}}{a_{1}}-1\right)\right]^{k-1}\frac{(-M_{0}(0)+g(t))}{a_{1}\left(\sqrt{2t}+a_{1}\right)}.\]
This means
\begin{equation}
\label{a}
\lim_{t\rightarrow\infty}\left[\left(\frac{\sqrt{2t}}{a_{1}}\right)^{k}-1\right]\left(\sqrt{2t}\right)^2=\frac{-M_{0}(0)}{2}(k).
\end{equation}
(b)
\begin{equation}
\label{b}
\sum_{i_{1},\cdots,i_{k};\prod_{j=1}^{k}i_{j}\not=1}i_{1}a_{i_{1}}a_{i_{2}}\cdots a_{i_{k}}\overline{a_{i_{1}+\cdots+i_{k}}}=O\left(\frac{1}{a_{1}^4}\right)=O\left(\frac{1}{(2t)^2}\right).
\end{equation}
Now we come back to estimate the right-hand side of (\ref{timerescaling}). By $(\ref{a}), (\ref{b})$ 
\begin{equation}
\label{123}
\left[\left(\frac{\sqrt{2t}}{a_{1}}\right)^{k}-1\right]\left[\sum_{i_{1},\cdots,i_{k};\prod_{j=1}^{k}i_{j}\not=1}i_{1}a_{i_{1}}a_{i_{2}}\cdots a_{i_{k}}\overline{a_{i_{1}+\cdots+i_{k}}}\right]=o\left(\frac{1}{2t}\right),
\end{equation}
and
\begin{equation}
\label{124}
\left[\sum_{i_{1},\cdots,i_{k};\prod_{j=1}^{k}i_{j}\not=1}i_{1}a_{i_{1}}a_{i_{2}}\cdots a_{i_{k}}\overline{a_{i_{1}+\cdots+i_{k}}}\right]=o\left(\frac{1}{2t}\right).
\end{equation}
By $(\ref{a})$,
\begin{equation}
\label{125}
\lim_{t\rightarrow\infty}M_{k-1}\left[\left(\frac{\sqrt{2t}}{a_{1}}\right)^{k}-1\right](2t)=-\frac{M_{0}(0)}{2}(k)M_{k-1}.
\end{equation}
We apply (\ref{123}), (\ref{124}) and (\ref{125}) to $(\ref{timerescaling})$ and obtain that for $k\geq 2$,
\[
\lim_{t\rightarrow\infty}\left[\overline{a_{k}}\left(\sqrt{2t}\right)^{k}-M_{k-1}\right]\left(\sqrt{2t}\right)^2=-\frac{M_{0}(0)}{2}(k)M_{k-1}.
\]
By taking complex conjugate of the above identity, we have for $k\geq 2$
\[
\lim_{t\rightarrow\infty}\left[a_{k}\left(\sqrt{2t}\right)^{k}-\overline{M_{k-1}}\right]\left(\sqrt{2t}\right)^2=-\frac{M_{0}(0)}{2}k\overline{M_{k-1}}.
\]
\end{proof}
\end{proof}

\subsection{Geometric characterization of a rescaling behavior for moving domains}
\label{sub4.3}
Assume $f_{n}(\xi,t)$ is a global strong degree $n\geq 2$ polynomial solution to (\ref{PG1}). We aim to give a geometric characterization for  a rescaling behavior of $\Omega(t)=f_{n}(D,t)$ as follows:
\begin{thm}
\label{geo}
Assume $f_{n}(\xi,t)$ is a global strong degree $n\geq 2$ polynomial solution to (\ref{PG1}) with Richardson complex moments $\{M_{k}\}_{k\geq 0}$. Let $\Omega(t)=f_{n}(D,t)$ and $\Omega'(t)=\{x|\sqrt{\mid\Omega(t)\mid/\pi}x\in\Omega(t)\}$ which has area $\pi$. Denote $n_{0}=\min\{k\geq 1\mid M_{k}\neq 0\}$. We show that
\begin{align}
\lim_{t\rightarrow\infty}\max_{z\in\partial\Omega'(t)}\left| \left| z\right|-1\right|(2t)^{1+\frac{n_{0}}{2}}&=\left| M_{n_{0}}\right|\neq 0,\notag\\
\lim_{t\rightarrow\infty}\max_{z\in\partial\Omega'(t)}\left| \kappa(t,z)-1\right| (2t)^{1+\frac{n_{0}}{2}}&=(n_{0}-1)(n_{0}+1)\left| M_{n_{0}}\right|\notag
\end{align}
where $\kappa(t,z)$ is the curvature for $z\in\partial\Omega'(t)$. This says that the decay rate $t^{-(1+n_{0}/2)}$ is the best rate we can get and the rescaling behaviors are precisely stated.
\end{thm}
\begin{proof}
Denote $f_{n}(\xi,t)=\sum_{i=1}^{n}a_{i}(t)\xi^{i}$. Let $g(\xi,t)$ be the Riemann mapping where $g:D\rightarrow\Omega'(t)$, $g^{'}(0,t)>0$ and $g(0,t)=0$, then
\[g(\xi,t)=\frac{f_{n}(\xi,t)}{\sqrt{2t+M_{0}(0)}}.\]
(a)By (\ref{rescaling100}), 
\[\lim_{t\rightarrow\infty}\left[ g(\xi,t)-\xi\right](2t)^{1+\frac{n_{0}}{2}}=\overline{M_{n_{0}}}\xi^{n_{0}+1}.\]
Therefore, 
\[\lim_{t\rightarrow\infty}\max_{z\in\partial\Omega'(t)}\left| \left| z\right|-1\right|(2t)^{1+\frac{n_{0}}{2}}=\left| M_{n_{0}}\right|\neq 0.\]
(b)Let  $A(t)=a_{1}(t)-\sqrt{2t+M_{0}(0)}$, then
\begin{align}
f_{n}(\xi,t)&=\left(\sqrt{2t+M_{0}(0)}\xi\right)+\left(\sum_{k=2}^{n}a_{k}(t)\xi^{k}\right)+A(t)\xi\notag\\
g(\xi,t)&=\xi+\left(\sum_{k=2}^{n}\frac{a_{k}(t)}{\sqrt{2t+M_{0}(0)}}\xi^{k}\right)+\left(\frac{A(t)}{\sqrt{2t+M_{0}(0)}}\xi\right).
\end{align}
Here $\kappa-1$ can be expressed by $g$ as follows:
\begin{equation}
\kappa-1=\frac{1}{\left| g^{'}\right|}Re\left(1+\frac{g^{''}\xi}{g^{'}}\right)-1 =\left(\frac{1}{\left| g^{'}\right|}-1\right)+\frac{1}{\left|g^{'}\right|}Re\left(\frac{g^{''}\xi}{g^{'}}\right).
\end{equation}
We set
\begin{align}
p=&\left(\sum_{k=1}^{n-1}\frac{(k+1)a_{k+1}}{\sqrt{2t+M_{0}(0)}}\xi^{k}\right)+\left(\frac{A(t)}{\sqrt{2t+M_{0}(0)}}\right)\notag\\
q=&\left(\sum_{k=2}^{n-1}\frac{(k-1)(k+1)a_{k+1}}{\sqrt{2t+M_{0}(0)}}\xi^{k}\right)-\left(\frac{A(t)}{\sqrt{2t+M_{0}(0)}}\right)\notag
\end{align}
and obtain  $g^{'}(\xi,t)=1+p$ and $g^{''}(\xi,t)\xi=p+q$. Therefore, we express $\kappa-1$ by $p, q$ as follows:
\begin{align}
\kappa-1=&\frac{Re(p)}{\left| g^{'}\right|}\left(\frac{2Re(p)+\left| p\right|^2}{\left(1+\left|g^{'}\right|\right)^2}\right)+\frac{Re(q)}{\left| g^{'}\right|}-\left(\frac{\left| p\right|^2}{\left| g^{'}\right|\left(1+\left|g^{'}\right|\right)}+\frac{1}{\left| g^{'}\right|}Re\left(p^2\right)\right)\notag\\
                              &+\frac{1}{\left| g^{'}\right|}Re\left(p^3-p^4+\cdots\right)+\frac{1}{\left| g^{'}\right|}Re\left(-qp+qp^2+\cdots\right).\notag                           
\end{align}
By applying Lemma~\ref{lemma4.2}, we can obtain the precise decay rates of $p$ and $q$  to calculate the precise decay rate of $|\kappa-1|$.  Here we treat two cases differently as follows:\\
(Case i)Assume that $M_{1}\neq 0$. That is $n_{0}=1$. The sharp decay rate of $p$ is $t^{-3/2}$.\\
$\bullet$ If $M_{2}\neq 0$, then the sharp decay rate of $q$  is $t^{-2}$. Therefore the sharp decay rate of $|\kappa-1|$ is $t^{-2}$.\\
$\bullet$ If $M_{2}=0$ and $M_{3}\neq 0$, then the sharp decay rate of $q$ is $t^{-5/2}$. Therefore the sharp decay rate of $|\kappa-1|$ is $t^{-5/2}$.\\
$\bullet$ If $M_{2}=0$, $M_{3}=0$ and $M_{4}\neq 0$, the sharp decay rate of $q$ is $t^{-3}$. Therefore, the sharp decay rate  of $|\kappa-1|$ is $t^{-3}$.\\
$\bullet$ Finally, if $M_{2}=M_{3}=M_{4}=0$, then the sharp decay rate of $q$ is $t^{-3}$. In this case, the sharp decay rate of $|\kappa-1|$ is $t^{-3}$.\\
In (Case i), it is clear that
 \[\lim_{t\rightarrow\infty}\max_{z\in\partial\Omega'(t)}\left|\kappa(t,z)-1\right|(2t)^{1+\frac{n_{0}}{2}}=0=(n_{0}-1)(n_{0}+1)\left| M_{n_{0}}\right|.\]
(Case ii)Assume $M_{1}=0$. That is $n_{0}=\min\{k\geq 1\mid M_{k}\neq 0\}\geq 2$. The sharp decay rates of $q$ and $p$ are both $t^{-(1+n_{0}/2)}$. In this case, the sharp decay rate of $|\kappa-1|$ is
\[\left|\kappa-1\right|=O\left(\frac{1}{t^{1+\frac{n_{0}}{2}}}\right).\]
Moreover, in (Case ii), 
\begin{align}
&\lim_{t\rightarrow\infty}\max_{z\in\partial\Omega'(t)}\left|\kappa(t,z)-1\right|(2t)^{1+\frac{n_{0}}{2}}\notag\\
=&\lim_{t\rightarrow\infty}\max_{\xi\in\partial D}\left|Re(q(\xi,t))\right|(2t)^{1+\frac{n_{0}}{2}}\notag\\
=&\lim_{t\rightarrow\infty}\left|\frac{(n_{0}-1)(n_{0}+1)a_{n_{0}+1}}{\sqrt{2t+M_{0}(0)}}\right|(2t)^{1+\frac{n_{0}}{2}}\notag\\
=& (n_{0}-1)(n_{0}+1)\left| M_{n_{0}}\right|.\notag
\end{align}

\end{proof}
\section{Rescaling behavior in Stokes flow}
\label{sec8}
Assume $\sum_{i=1}^{n}a_{i}(t)\xi^{i}$ is a global polynomial solution to zero surface tension Stoke flow driven by injection with strength $2\pi$. The invariant quantities $C_{k}=\sum_{j=1}^{n-k}ja_{j}\overline{a_{j+k}}$ for $1\leq k\leq n-1$ as shown in Cummings, Howison and King~\cite{cummings}. Here $C_{0}(t)=\sum_{j=1}^{n}ja_{j}(t)\overline{a_{j}(t)}=C_{0}(0)+2t$. Since $|a_{j}(t)|\leq ja_{1}(t)$, then 
\[C_{0}(0)+2t=\sum_{j=1}^{n}ja_{j}(t)\overline{a_{j}(t)}\leq\sum_{j=1}^{n}j^3a_{1}^2(t)\]
which implies that $a_{1}^2(t)\geq\frac{1}{\sum_{j=1}^{n}j^3}(2t+C_{0}(0))$. 
\subsection{Decay of coefficients in terms of invariant quantities}
\begin{lemma}
\label{bird}
If $f_{n}(\xi,t)=\sum_{i=1}^{n}a_{i}\xi^{i}$ is a global degree $n\geq 2$ polynomial solution to (\ref{stokes1}), then $\overline{a_{k}(t)}a_{1}(t), k\geq 2$ is bounded.
\end{lemma}
\begin{proof}
It is sufficient to prove that $\overline{a_{n-j}}a_{1}$ is bounded for $0\leq j\leq n-2$.\par
Since $C_{n-1}=a_{1}\overline{a_{n}}$. Therefore the above result holds for  $j=0$. Now assume that
$\overline{a_{n-j}}a_{1}$ is bounded for $0\leq j\leq s_{0}$.\par
Since
\[C_{n-(s_{0}+2)}=a_{1}\overline{a_{1+(n-s_{0}-2)}}+\sum_{j=2}^{s_{0}+2}ja_{j}\overline{a_{j+(n-s_{0}-2)}},\]
therefore, 
\begin{equation}
\label{aa}
a_{1}\overline{a_{n-(s_{0}+1)}}=C_{n-(s_{0}+2)}-\sum_{j=2}^{s_{0}+2}ja_{j}\overline{a_{j+(n-s_{0}-2)}}.
\end{equation}
Here, for $2\leq j\leq s_{0}+2$,
\[\left|a_{j}\overline{a_{j+(n-s_{0}-2)}}\right|=\left|\left(\frac{a_{j}}{a_{1}}\right)\left(a_{1}\overline{a_{j+(n-s_{0}-2)}}\right)\right|\leq j\left|a_{1}\overline{a_{j+(n-s_{0}-2)}}\right|\]
which is bounded since we assume $\overline{a_{n-j}}a_{1}$ is bounded for $0\leq j\leq s_{0}$. Therefore, $\sum_{j=2}^{s_{0}+2}ja_{j}\overline{a_{j+(n-s_{0}-2)}}$ is bounded. We obtain that $a_{1}\overline{a_{n-(s_{0}+1)}}$ in (\ref{aa}) is bounded.
\end{proof}
By (\ref{bird}) and similar argument as that of Hele-Shaw flows, we obtain the following results.
\begin{lemma}
If $f_{n}(\xi,t)=\sum_{i=1}^{n}a_{i}\xi^{i}$ is a global degree $n\geq 2$ polynomial solution to (\ref{stokes1}), then for $2\leq k\leq n$, 
\[\lim_{t\rightarrow\infty}\left[\overline{a_{k}}a_{1}-C_{k-1}\right](\sqrt{2t})^2=\sum_{j=2}^{n-(k-1)}jC_{j-1}\overline{C_{j+k-2}}\]
and
\[\lim_{t\rightarrow\infty}\left[a_{1}(t)-\sqrt{2t+C_{0}(0)}\right](\sqrt{2t})^{3}=-\frac{1}{2}\sum_{k=2}^{n}k|C_{k-1}|^2.\]
\end{lemma}
\begin{thm}
If $f_{n}(\xi,t)=\sum_{i=1}^{n}a_{i}\xi^{i}$ is a global degree $n\geq 2$ polynomial solution to (\ref{stokes1}), then
\[\lim_{t\rightarrow\infty}\left[f_{n}(\xi,t)-\sqrt{2t+C_{0}(0)}\xi\right]\sqrt{2t}=\sum_{k=2}^{n}\overline{C_{k-1}}\xi^{k}.\]

\end{thm}
\subsection{Geometric characterization of a rescaling behavior for moving domains}
Given $f_{n}(\xi,t)$ which is a global  degree $n\geq 2$ polynomial solution  to (\ref{stokes1}). We aim to give a geometric characterization for  a rescaling behavior of $\Omega(t)=f_{n}(D,t)$ as follows:
\begin{thm}
Assume $f_{n}(\xi,t)$ is a global degree $n\geq 2$ polynomial solution to (\ref{stokes1}) with invariant quantities  $\{C_{k}\}_{k\geq 1}$. Let $\Omega(t)=f_{n}(D,t)$ and $\Omega'(t)=\{x|\sqrt{\mid\Omega(t)\mid/\pi}x\in\Omega(t)\}$ which has area $\pi$. Then 
\begin{align}
\lim_{t\rightarrow\infty}\max_{z\in\partial\Omega'(t)}\left| \left| z\right|-1\right|(2t)&=\max_{\xi\in\partial D}\left|\sum_{k=2}^{n}\overline{C_{k-1}}\xi^{k}\right|\neq 0,\notag\\
\lim_{t\rightarrow\infty}\max_{z\in\partial\Omega'(t)}\left| \kappa(t,z)-1\right| (2t)&=\max_{\xi\in\partial D}\left|\sum_{k=1}^{n-1}(k-1)(k+1)\overline{C_{k}}\xi^{k}\right|.\notag
\end{align}
where $\kappa(t,z)$ is the curvature for $z\in\partial\Omega'(t)$.\end{thm}
\begin{proof}
The proof is similar to that of  Theorem~\ref{geo}.
\end{proof}

\pagebreak
\section*{Acknowledgements}
This paper is an extraction from my Ph. D thesis. I am indebted to my advisor, Govind Menon, for many things, including his constant guidance and important opinions. This material is based upon work supported by the National Science
Foundation under grant nos. DMS 06-05006 and DMS 07-48482.


\bibliography{main0}

\end{document}